\documentclass[twocolumn]{aastex631} 

\usepackage{amsmath}

\begin{document}

\title{A Butterfly's Eye Camera for Intensity Interferometry with Cherenkov Telescopes}

\correspondingauthor{J. Cortina}

\author[0000-0003-4576-0452]{J. Cortina}
\affiliation{Centro de Investigaciones Energ\'eticas, Medioambientales y Tecnol\'ogicas, E-28040 Madrid, Spain}
\affiliation{Instituto de Astrof\'{\i}sica de Canarias, E-38205 La Laguna, Tenerife, Spain}
\email{juan.cortina@ciemat.es}

\author[0000-0003-1033-5296]{A. Cifuentes Santos}
\affiliation{Centro de Investigaciones Energ\'eticas, Medioambientales y Tecnol\'ogicas, E-28040 Madrid, Spain}

\author[0000-0002-4758-9196]{T. Hassan}
\affiliation{Centro de Investigaciones Energ\'eticas, Medioambientales y Tecnol\'ogicas, E-28040 Madrid, Spain}

\author[0009-0004-5848-8763]{F. Fr\'{\i}as}
\affiliation{Departamento de Astrof\'{\i}sica, Universidad de La Laguna, E-38205 La Laguna, Tenerife, Spain}
\affiliation{Centro de Investigaciones Energ\'eticas, Medioambientales y Tecnol\'ogicas, E-28040 Madrid, Spain}





\begin{abstract}

In recent years, imaging atmospheric Cherenkov telescopes (IACTs) have emerged as promising platforms for optical interferometry through the use of intensity interferometry. IACTs combine large segmented mirrors, photodetectors with nanosecond-scale time response capable of detecting signals from just a few photo-electrons, and array configurations with baselines of hundreds of meters. As a result, all major IACT facilities have now been upgraded to function also as optical intensity interferometers, achieving sensitivities an order of magnitude better than their predecessor, the Narrabri Stellar Intensity Interferometer.
However, further improvements in sensitivity are currently limited by key IACT design constraints, namely the combination of poor optical quality and small focal ratios.
Here we present three practical implementations of the 'I3T concept', in which segments of the IACT primary mirror are focused onto different pixels of its camera. This approach yields several unexpected but significant advantages. Optics with larger focal ratios allow to integrate narrow-band optical filters, while lower photon fluxes enable to deploy next-generation photodetectors operating in photon-counting mode. 
We demonstrate that this so-called 'Butterfly’s Eye' configuration enhances the sensitivity of IACT-based intensity interferometers by a factor between 4 and 6. Moreover, as originally envisioned, the I3T design introduces imaging capabilities on angular scales from 2 to 40 milliarcseconds, unlocking new scientific opportunities such as direct surface imaging of nearby red giants. Besides, realistic simulations show that it can have a transformative impact on at least two key science cases: imaging the earliest stages of nova ejecta, and measuring the oblateness and circumstellar disks of fast-rotating stars.

\end{abstract}



\section{Introduction}

Intensity interferometry has experienced a revival in astrophysics, driven by advances in high-speed photodetectors and the unique capabilities of imaging atmospheric Cherenkov telescopes (IACTs). Originally pioneered by Robert Hanbury Brown and Richard Twiss in the 1950s \citep{HB1974}, to measure stellar diameters through the correlation of intensity fluctuations from independent telescopes, intensity interferometry lost prominence due to the rise of amplitude interferometry, which offered higher sensitivity. However, the technique has reemerged because IACTs, such as VERITAS \citep{VERITAS}, MAGIC \citep{magic_2019} or H.E.S.S. \citep{HESS}, even if built for Very High Energy (VHE) $\gamma$-ray observations \citep{handbook_iacts}, happen to fulfill many of the requirements of an intensity interferometer at visible wavelengths.

Recent studies demonstrate the potential of intensity interferometry for resolving stellar features and binary star systems \citep{Dravins, cta}, particularly in the optical regime where amplitude interferometry faces challenges to control the phase over long baselines \citep{Eisenhauer}. 
By utilizing arrays of Cherenkov telescopes for intensity interferometry, astronomers can achieve sub-milliarcsecond angular resolution.

The so-called 'I3T concept' \citep{I3T}  represents a new approach to using IACTs: aperture synthesis techniques are applied to pairwise and triple correlation of sub-pupil intensities in order to reconstruct the image of a celestial source. In practical terms, the I3T concept allows an individual IACT to reach diffraction-limited imaging of celestial sources. In this work, we will evaluate the feasibility of a specific implementation of the concept in the MAGIC telescopes. The implementation will also illustrate that the concept brings an additional bonus because it allows to use faster photodetectors and simpler secondary optics. It should be noted that an alternative implementation of the I3T concept has been proposed in \citep{nolan_pupil}.

\section{Current implementations in IACTs: the case of MAGIC and LST}
\label{sec:current_implementation}

The implementation of the intensity interferometry technique in IACTs is conceptually similar. We will deal here with two specific ones, MAGIC and LST, which may allow an easier technical implementation of the I3T concept because they are equipped with active optics. They should anyway be considered as case studies.

MAGIC is a system of two IACTs located at the Roque de los Muchachos Observatory (ORM) on the island of La Palma in Spain \citep{magic_upgrade1}. Equipped with 17~m diameter parabolic reflectors and photomultiplier (PMT) cameras, the telescopes record images of extensive air showers initiated by $\gamma$-rays with energies of few tens of GeV \citep{magic_upgrade2}. After initial exploratory measurements in April 2019 \citep{magic_2019}, a new easy-to-deploy instrumental setup with essentially 100\% duty cycle has been installed and is currently operational. See \cite{spie2022} and \cite{magic_2024} for a full description and a performance evaluation. Here we will only describe the setup parameters that are relevant for this study.

The MAGIC reflector follows a parabolic shape to minimize the time spread at the focal plane. The focal length is 17~m. The reflector is formed by $\sim$250 1~m$^2$ spherical mirror tiles. Because the mirrors are supported by a light-weight space-frame, the reflector shape deforms with elevation and each tile is equipped with two actuators to compensate for this deformation. This so-called Active Mirror Control (AMC) adjusts the mirror in a few seconds. Approximately 70\% of the light from a point source reflected in the mirror is focused inside a pixel.

We strongly profit from the AMC during intensity interferometry observations. On one hand, the reflectors are not focused to stars at infinity during VHE observations but we can change focus to infinity in a matter of seconds before switching to interferometry.

%
On the other one, the AMC software can focus the entire reflector or arbitrary groups of mirror tiles to arbitrary positions in the camera focal plane (within a circle of roughly 1 deg radius):
\begin{itemize}
\item In its simplest configuration the entire reflector is focused on a pixel that is 24~cm from the camera center.
\item  In the so-called chessboard configuration half of the mirror tiles (arranged like the black squares in the chessboard) are focused to a pixel while the other half (the white squares) are focused on a second pixel. This allows to study the correlation between the 'white' and 'black' mirrors, which corresponds to a distribution of baselines always $<$17~m, providing an almost direct way of measuring the zero-baseline correlation.
\item  In other configurations, two roundish groups of 12-21 tiles are focused to each of the two pixels. This was inspired by the I3T concept \citep{I3T} and allows to measure the correlation for a specific baseline $<$17~m and different orientations in the sky.
\end{itemize}

Each telescope is equipped with a 1039-pixel PMT camera at the primary focus. The PMTs are 25.4~mm diameter and reach a peak quantum efficiency (QE) of 33\% at a wavelength around 400 nm. A light concentrator is mounted on top on each PMT. The distance between PMT centers is 30~mm. 

As we shall see the signal to noise of the correlation of the telescope signals is insensitive to the width of the optical passband of the detected light. However, a filter spectral response with sharp cutoffs at the edges of the passband improves the sensitivity of the measurement, and, even more importantly, the bright stars in typical interferometry observations would damage the PMTs after a short exposure time. Therefore, interferometry observations require installing filters in front of the pixels connected to the correlator. The filters in MAGIC are centered at 425 nm and have a Full Width Half Maximum (FWHM) of 25~nm. Each of them is placed in front of a PMT. Several pairs of filter and PMTs are in use, so as to allow the above-described observations modes.

The electrical signals of the PMTs are amplified (AC coupled) and then transmitted to a separate  readout location. The PMT pulses have a time response of 2.5~ns. A digitizer samples the PMT waveform with 500 MS/s. The bandwidth of the acquisition chain is limited by the digitizer to $\sim$130 MHz. The digitized waveforms are correlated realtime with a GPU-based system. For more details see \cite{magic_2024}.

The Cherenkov Telescope Array Observatory CTAO \citep{cta} will consist of two arrays of IACTs at the northern and southern hemispheres. With its large number of telescopes and different baselines, equipping CTAO for intensity interferometry would deliver a significant increase in performance respect to the current IACT arrays (see for instance section 14.3 of \cite{cta} for some of its scientific goals). The primary mirrors of the CTAO IACTs will be of three different sizes, optimized for different $\gamma$-ray energy ranges. The proposed sub-arrays of four Large-Sized Telescopes (LST \citep{lst}) at CTAO-North and CTAO-South should be equipped with the largest mirror area (roughly 400~m$^2$). CTAO will also be equipped with tens of Medium-Sized Telescopes (MST, 12 m diameter) and Small-Sized Telescopes (SST, $\sim$4 m diameter).

A full-size prototype (LST-1) was installed at the CTAO-North site at ORM in 2018, at a distance of around 100~m from the MAGIC telescopes, is completing its software integration and scientific validation. The installation of three more LSTs should be complete in 2026. The design of the LST was inspired by MAGIC so many of its parameters are similar. The reflector is equally tessellated (198 hexagonal 2~m$^2$ mirror tiles) and parabolic in shape but has a larger diameter, 23~m, and a larger f/D=1.2. The telescope is also equipped with an AMC. The camera has 1855 PMTs with a peak QE of 42\%. The distance between PMTs is 50~mm. The PMTs of MAGIC and LST have a similar time response. An instrumental setup similar to the one operational in MAGIC was installed in 2024 and allows interferometry observations with LST-1 and the two MAGIC telescopes (see \cite{spie2022, spie_2024} for details). It is equipped with the same interference filters as MAGIC. It may be extended to all four LSTs at ORM. 

Compared to the current MAGIC interferometer, the combined array of MAGIC and the four LSTs is expected to boost the sensitivity by roughly a factor 16 \citep{spie2022, magic_2024}.

\section{Limitations to increasing the performance of an IACT-based interferometer}

Equation 5.17 in \cite{HB1974} allows to calculate the significance (signal over noise) of the correlation for a given 
experimental setup and unpolarized light. The equation can be written as: 
\begin{equation}
\begin{split}
S/N =\; & A~\alpha ~q ~n ~|V(d)|^2 ~\sqrt{b_\nu} ~F^{-1} ~\sqrt{T/2} ~(1 + a)^{-1} ~\sigma
\end{split}
\label{eq:significance}
\end{equation}
where A is the telescope mirror area, $\alpha$ is the QE of the PMTs for the filter's central wavelength, q is the efficiency of the remaining optics, n is the star's differential photon flux, $b_v$ is the effective cross-correlation  electrical bandwidth, $F$ is the excess noise factor of the photodetector and $T$ is the observation time, and $(1+a)$ allows for the presence of Night Sky Background (NSB), stray light and dark current: a is the average background to starlight ratio. Finally $\sigma$ is the normalized spectral distribution of the light after the filter as defined in formula (5.6) of \cite{HB1974}. $\sigma$ would be equal to 1 if the filter transmission curve is a boxcar function and the spectrum of the light is flat. We assume that the visibility does not change and the source spectrum is flat over the filter bandwidth. 

When dealing with an array of N$_{\mathrm{tel}}$ equal telescopes with redundant baselines the S/N increases with the square root of the telescope pairs. The number of telescope pairs is N$_{\mathrm{tel}}(\mathrm{N}_{\mathrm{tel}}-1)/2$ so the array's S/N increases roughly linearly with N$_{\mathrm{tel}}$.

\subsection{Night sky background}

Equation \ref{eq:significance} shows that NSB effectively prevents observations of stars producing a photon flux less than the NSB collected over the FOV of a camera pixel. The smallest $\gamma$-ray images recorded at an IACT camera are a few arcminutes in angular size so IACTs have no incentive to improve their optical point spread function (PSF) and correspondingly their pixel FOV below $\sim$0.05$^{\circ}$. Both MAGIC and LST pixels have FOVs of 0.1$^{\circ}$. This FOV corresponds, under dark-sky conditions, to a star with B$\sim$8.5$^\mathrm{m}$. Fields closer to the galactic plane or higher zenith angles bring the limiting magnitude to 8$^\mathrm{m}$ or even 7$^\mathrm{m}$. In fact, MAGIC currently observes in interferometry mode during full moon nights or in the 2-3 nights around them, effectively limiting viable targets to the $\sim$4.5 - 5$^\mathrm{m}$ range.

One may consider to reduce the FOV of the pixel, using for instance a diaphragm. This is possible for LST, where the optical PSF has a size (diameter containing 80\% of light) better than 35 mm at the center of the camera, which is smaller than the pixel size of 50 mm. Even if there may be some losses due to tracking precision, a diaphragm would allow reducing the FOV by a factor 2, increasing the limiting magnitude to almost 10$^\mathrm{m}$. Using a diaphragm in MAGIC is more challenging because the PSF is essentially as large as the pixel, so the limiting magnitude of MAGIC remains at B$\sim$9$^\mathrm{m}$.

\subsection{Photodetectors}
\label{subsec:photodetectors}

As already introduced in section \ref{sec:current_implementation}, the PMTs currently equipped both in MAGIC and LST1 have similar QEs (33 and 42\% for MAGIC and LSTs respectively) and time response (pulses of 2.5 ns FWHM). Following equation \ref{eq:significance}, one can increase the sensitivity of an intensity interferometer by increasing the QE or the bandwidth. In fact, triggered by applications to Quantum Key Distribution, fluorescence microscopy or time of flight Positron Emission Tomography, a good number of photodetector technologies now provide time resolutions well below 1~ns with an acceptable QE (see \cite{fast_photodetectors} for a recent review). These technologies include conventional PMTs, single photon avalanche diodes (SPADs), hybrid photodetectors (HPDs), Silicon photomultipliers (SiPMs), Microchannel Plate PMTs (MCP-PMTs) and superconducting nanowire single photon detectors (SNSPDs). Several parameters are relevant to characterize their performance: 
\begin{itemize}
\item Transit Time Spread (TTS) describes the statistical variation in the arrival times of the electrical pulses produced when identical photons hit a photodetector. As above-mentioned, the PMTs in current IACTs have a TTS (FWHM) around 2 ns. Faster PMTs may reach a TTS FWHM of hundreds of ps, HPDs and MCP-PMTs few tens of ps, whereas SPADs and SNSPDs reach $<$15 ps.
\item QE, or more precisely Photon Detection Efficiency (PDE), is the ratio of detected photoelectrons to incident photons. PDE changes with wavelength of the incident photon. Peak PDE ranges from 10\% to 60\% in some SiPMs. 
\item Single photon charge resolution is the ability of a photodetector to generate a unique response to a single photon. It was represented by F in equation \ref{eq:significance}. Whereas PMTs show poor single photon charge resolution (F=1.15 or more), SiPM, SPADs, MCP-PMTs and HPDs have $F$ close to unity (although some of them suffer from crosstalk). 
\item Dark count rate (DCR) refers to the rate of photoelectrons in a photodetector operated under a totally dark condition. In general, PMTs, HPDs, MCP-PMTs have low dark currents below 1 kHz in the full device, while
  DCR in SiPM or SPADs typically exceeds 50 kHz~mm$^{-2}$.
\item Afterpulses are spurious pulses that appear after the initial photoelectron. SPADs and SiPMs experience a high afterpulse rate. HPDs and SNSPDs are essentially free of afterpulses.
\item Dead time: some of these photodetectors remain inoperative for some time after a photoelectron. For instance, individual cells of SPADs or SiPM only allow photoelectron detection rates of MHz. PMTs have no dead time.  
\item The sensitive area can range from a few $\mathrm{\mu}$m for SNSPD or the individual cell of a SPAD or SiPM to several mm in HPDs and even cm in MCP-PMTs or PMTs.
\item Operating temperature: PMTs, MCP-PMTs and HPDs operate at room temperature, whereas SiPM or SPADs show high DCR unless cooled to at least -20 C. SNSPDs operate below 4 K.
\end{itemize}

Some of these photodetectors are difficult to apply to IACTs because IACT pixels have a size of a few cm. SNSPDs simply do not exist with this size. SPADs are typically smaller than 1 mm. SiPM can reach sizes of cm but their time resolution degrades quickly. Besides, the DCR in both SPADs and SiPM scales with sensitive area. In general, it is difficult to find photodetectors with time resolutions better than 100 ps combined with photosensitive areas larger than 1 cm$^2$.

\begin{figure}
  \begin{center}
      \includegraphics[width=0.52\textwidth]{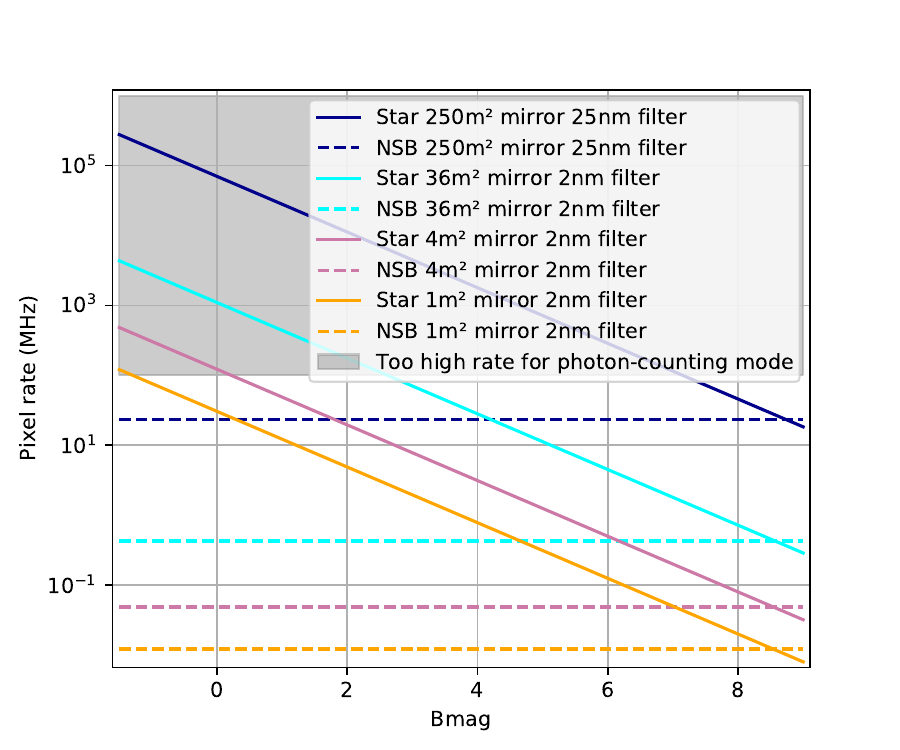}
  \end{center}
  \caption{ \label{fig:Phe_rates}
            Photoelectron rates expected in a pixel as a function of star's B magnitude. The uppermost solid line corresponds to the current MAGIC telescope, that is, full reflector focused into the pixel, 25 nm filter passband and the PDE of the current photomultipliers. The other three solid lines correspond to the three Butterfly configurations that are described in the text, all with a narrower filter passband of 2 nm and enhanced PDE but with three different submirror areas. The NSB rate under dark sky conditions for the four configurations is also drawn as a reference. We have shadowed the area where the rate is too high to operate in photon-counting mode (i.e. rate above 100 MHz).} 
\end{figure}

Fast photodetectors are also difficult to integrate with the correlators that are used in current IACTs. These are based on digitizers with sampling rates around 1 GSps, which are too slow for photodetectors with time resolutions below 1~ns. No digitizer in the market can sustain a sampling rate exceeding 10 GSps, as would be necessary for time resolutions $<$100~ps.

In fact, the above-mentioned applications of fast photodetectors do not sample the analog waveform. They rather discriminate pulses generated by individual photons and generate digital pulses. In other words, they operate in 'photon counting mode'. For an intensity interferometer, these digital pulses may be fed into a correlator device to look for time coincidences with other pixels or time tag them using a reference common clock for later correlation using GPUs or FPGAs. The photon counting mode is the only option to increase the bandwidth of the interferometer. The mode has in fact been tested successfully for intensity interferometry \citep{Guerin}.

This mode has another significant advantage. When correlating waveforms any correlated noise between the two channels, even with an amplitude significantly smaller than the amplitude of one photoelectron, will produce a signal that grows with time and will eventually limit the sensitivity of the correlated signal. This effect has been clearly observed when correlating signals from PMTs from the same MAGIC camera. Taking into account that the amplitude of correlation signals from stars for the MAGIC setup are typically on the order of 10$^{\mathrm{-6}}$, very small correlated noise components will dominate over these signals. This is not the case in the photon counting regime because the amplitude of each single photoelectron pulse is well above the noise. Once the photon has been discriminated, the noise will not accumulate in time. Other sub-dominant noise sources may be present, such as non-linearities in the time to digital converters, cross-talk between detectors, etc.

However, working in this mode introduces new challenges. Dead time becomes an issue for some  of the photodetectors, which may remain inactive for $>$1~ns after a photon detection. Even if the time resolution is below 1 ns, the shape produced by a single phe is typically long (ns scales), so single photon pulses start to overlap with rates in excess of hundreds of MHz (typically referred to as 'pile-up'). Effectively the bandwidth of the system becomes dominated by the shape of the single photon pulses. 

Hundreds of MHz is actually rather low for the typical rates detected in current IACTs for bright stars. The first set of lines in figure \ref{fig:Phe_rates} show the rates that are expected in one of the current MAGIC pixels for the parameters in \citep{magic_2019}. A B$<$3 star generates a rate well above 1~GHz. These high pixel rates result from a combination of the large mirror collection area and the relatively broad bandpass filter. Even a 7$^\mathrm{m}$ star, which is the expected limit magnitude for the combined array of MAGIC and the four LSTs, would produce a rate above 100 MHz.

\subsection{Narrow spectral channels and multi-spectral channel observations}

On the other hand, equation \ref{eq:significance} is notable in the fact that the S/N does not depend on the width of the optical passband. This means that one could reduce the passband to match the width of emission spectral lines in order to study the region of the star that is responsible for the line emission. Spectral lines in the expanding shells of novas or supernovas may be as wide as $\sim$10 nm but in most of the cases they are in the order of 1 nm in winds or decretion disks, and even narrower in other objects. 

One can also split the light into several spectral channels N$_{\mathrm{spectral}}$ of a much smaller passband and each of these channels would produce a signal with the same S/N. By combining all the spectral channels the total S/N of the interferometer increases by a factor $\sqrt{\mathrm{N_{spectral}}}$. However, both reducing the passband of a spectral channel with filters and splitting light in a large number of narrow spectral channels is extremely challenging with the current optics of IACTs.

\subsection{Effect of the optics of IACTs}

IACT reflectors have focal ratios (f/D) between 1 and 1.3 so light enters the pixel with a large range of incident angles going up to $\sim$25 deg. The spectral transmission curve of interference filters shifts to shorter wavelengths with incident angle. In the case of MAGIC, for instance, the original transmission curve has a boxcar shape with a FWHM of 20 nm but the effective transmission curve for light coming uniformly from the reflector develops two 'tails', which actually degrade the sensitivity by roughly 15\% (through the $\sigma$ parameter in equation \ref{eq:significance}). A filter with a narrower passband would result in an even stronger degradation of sensitivity so IACTs are currently restricted to 10-20 nm passbands. 

The same applies to the use of diffraction gratings or prisms to resolve the light into many spectral channels. There is no practical optical element that is efficient with such a low f/D and large PSF. 

Yet another negative aspect in the optics of an IACT comes from the fact that a time resolution better than 1 ns is generally not needed as the intrinsic time spread of shower photons is already larger \cite{Chitnis}. That is why the optics of most of the current IACTs follow a Davies-Cotton optical design, where the overall shape of the reflector is spherical. The time spread of a 10~m diameter spherical reflector is larger than 1~ns. 

Notable exceptions are the reflectors of MAGIC and LST, which follow a parabolic shape. The photon time distribution in MAGIC and LST has a width below $\sim$1 ns when focusing on-axis \citep{ph_m_garczarczyk, LST_requirements}. In these telescopes the limitation comes from the mechanical precision in the alignment of the tiles. An additional time spread is introduced if we focus the light far from the optical axis. As we shall see in the next section, both effects may increase the time spread to hundreds of ps. For the specific case of the first MAGIC telescope, it must be noted that the time spread is even larger because the tiles are staggered in two levels due to a design flaw.

Consequently, photodetectors with an RMS of a time resolution better than 100 ps would not improve the sensitivity of the interferometer. 

\section{Overcoming these limitations with the I3T concept}

The I3T concept was proposed as a way to turn an IACT into a diffraction-limited telescope of the same diameter \citep{I3T}. As explained by the proposers, such an instrument would have a significant scientific impact: imaging the surface of Betelgeuse, even in a coarse manner, would help to understand the processes occurring in the atmosphere of the late stages of red supergiants about to become a supernova, constraining its mass loss and convective instabilities. The authors also point to other significant science cases such as imaging of AGB and post-AGB stars, Be and B[e] stars circumstellar envelopes, winds, and even point to studies of extragalactic objects.

In the next sections we will describe a practical implementation of the concept to MAGIC and, even more importantly, we will show that the same concept also represents a way to overcome the above-mentioned limitations that prevent to increase the sensitivity and scientific reach of an IACT interferometer. In this way, this implementation also allows to increase the sensitivity of arrays of IACTs with baselines of hundreds of meters or even km. 

\begin{figure}
  \begin{center}
    \includegraphics[width=0.48\textwidth]{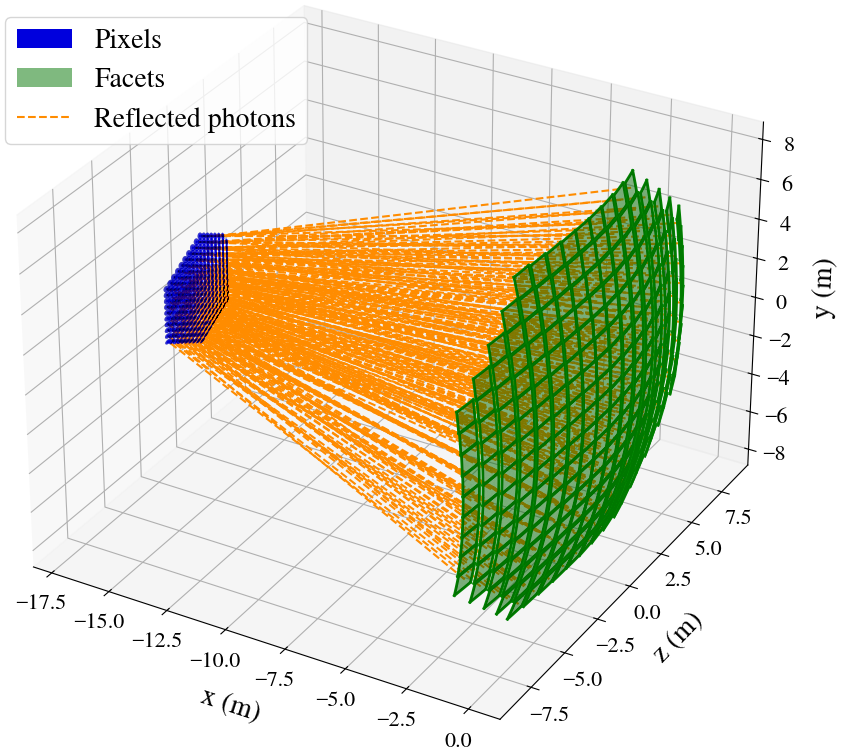}
  \end{center}
  \caption{
    \label{fig:Illustration_Butterflys_eye}
    An illustration of the I3T concept applied to a MAGIC IACT.
    Each of the tiles of the tesselated 17 m diameter primary mirror is focused to a
    pixel. In this figure, the pixels are spread over a very large area to better 
    illustrate the concept. As we shall see practical implementations do not require 
    such a large camera.}
\end{figure}
Figure \ref{fig:Illustration_Butterflys_eye} illustrates how the I3T concept could be implemented on one of the MAGIC telescopes. Each of the mirror tiles is focused onto an individual pixel. This already reduces the problem of pile-up because each photodetector collects light from a limited fraction of the reflector, which we will refer to as 'submirror'. However, it also represents a real bonus in terms of optical quality. For the most extreme case of a submirror of 1~m$^2$ f/D would increase to 17 in MAGIC (f= 17~m) and 28 in LST (f= 28~m). Assuming that each pixel is aligned to the corresponding submirror, light would enter the pixel with an angle $<$1.7$^{\circ}$ in MAGIC and LST. As a consequence, one may:
\begin{itemize}
\item Use a narrow interference filter down to $\sim$0.5 nm FWHM with no significant degradation in spectral response. 
\item Alternatively, use other readily available optical elements such as diffraction gratings or beamsplitters with a size of $\sim$30 mm.
\item Use a lens with a small diameter $\sim$30 mm to condense the light in a spot of $<$3 mm (assuming f= 17~m). 
\end{itemize}
As we shall see, this narrow transmission passband results in a further reduction in the photodetection rates and allows us to work in the photon counting regime.

\section{Butterfly optical configurations and expected photon arrival time distributions}

\subsection{Optical configurations}

Let us contemplate three optical configurations of a 'MAGIC Butterfly' telescope:
\begin{itemize}
    \item {\bf Butterfly 1}: The reflector is divided into 250 submirrors. Each of the submirror corresponds to a 1~m$^2$ tile and is focused into a different pixel. This will require a camera of 250 pixels. Since each pixel, including the corresponding focusing optics (a lens and possibly a baffle) will have a width similar to the optical PSF (30 mm), the camera will be relatively large, with a physical radius of \mbox{$\sim$25 cm.} 
    %
    \item {\bf Butterfly 2}: The reflector is divided into 64 submirrors. Each submirror consists of four adjacent 1~m$^2$ tiles, focused into a camera with 64 pixels. If each pixel had a size of 30 mm, the camera would have a radius of $\sim$15 cm. However, we will stay on the conservative side and assume that each pixel has 50 mm diameter, so the camera radius will be 25 cm.
    \item {\bf Butterfly 3}: The reflector is divided into only 7 submirrors. Each of them corresponds to 36~m$^2$ tiles.
    The camera will have 7 pixels. Even assuming that each pixel will need to be larger (because a larger $\sim$50~mm diameter lens will be needed to focus the light), the camera will have a radius of $<$10 cm. 
\end{itemize}

\begin{figure*}
  \begin{center}
    \includegraphics[width=0.85\textwidth]{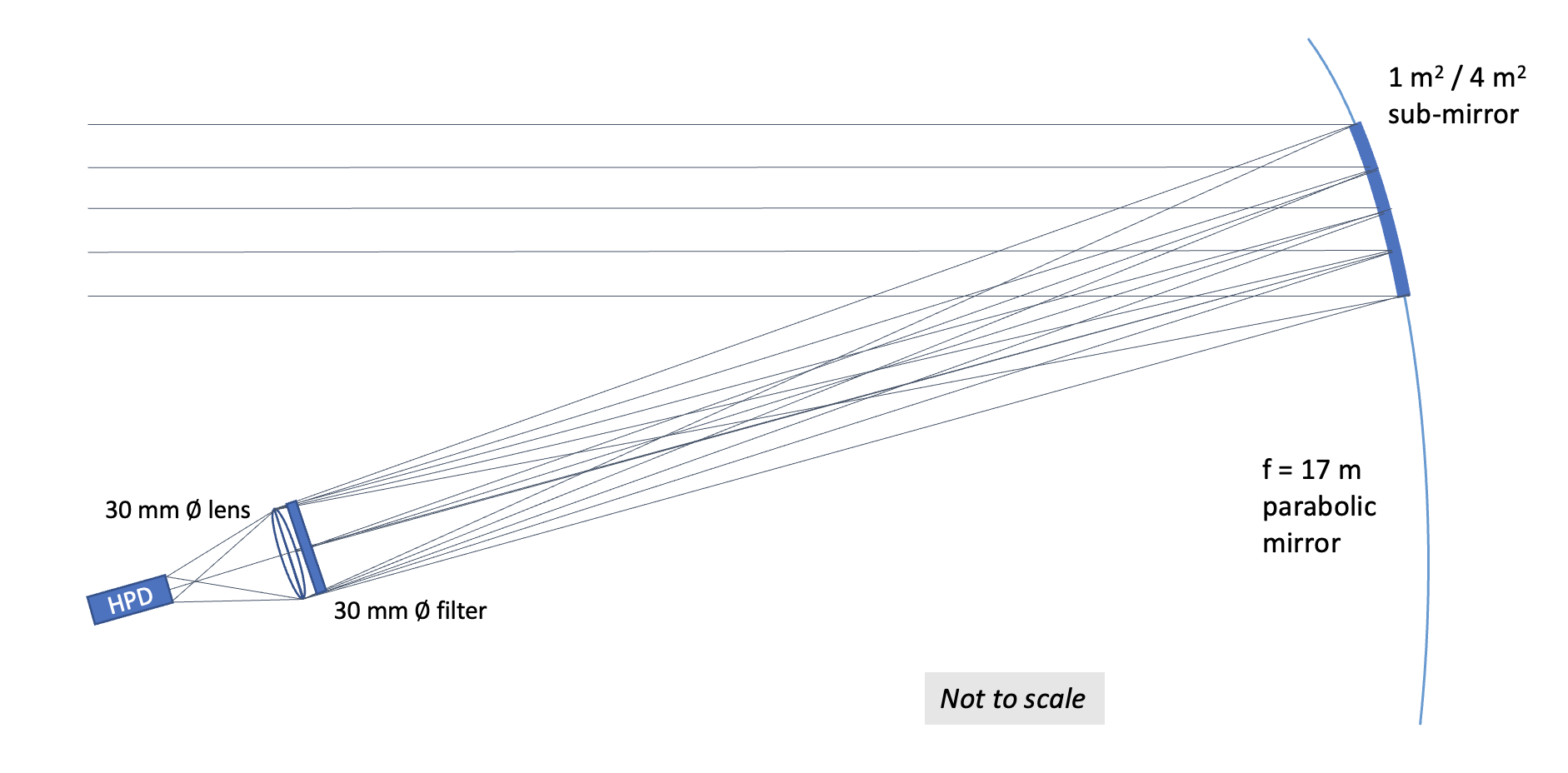}
  \end{center}
  \caption{
    \label{fig:scheme_optics}
    A very simplified scheme of the optics: light from a 1 m$^2$ or 4 m$^2$ section of the MAGIC reflector ('submirror', corresponding to 1 or 4 mirror tiles) is focused into a pixel which has been aligned in the direction of the submirror. A 30 mm diameter lens is employed to condense the light into a photodetector. Note that light coming from the sub-mirror covers essentially the whole lens due to the large PSF.}
\end{figure*}
Figure \ref{fig:scheme_optics} illustrates the optics of the first two configurations. The 1 or 4 m$^2$ submirror is focused into a pixel using the active mirror control. The elements of the pixel are aligned with the submirror. The first element is a 30 mm diameter filter. The filter is immediately followed by a converging lens, which focuses the light into the photosensitive area of an HPD. This is essentially a concept: a practical implementation may require more complex optics.

As all pixels would be aligned to a tile, the global shape of the camera would follow a roughly spherical shape. The setup resembles the eye of a butterfly.

The different configurations perform differently in terms of complexity and performance. A larger number of pixels and submirrors increases cost and complexity but, as we shall see, it results in improved time resolution and consequently sensitivity, as well as better coverage in the uv plane.

%

\begin{figure*}
	\centering
        \includegraphics[width=0.92\textwidth]{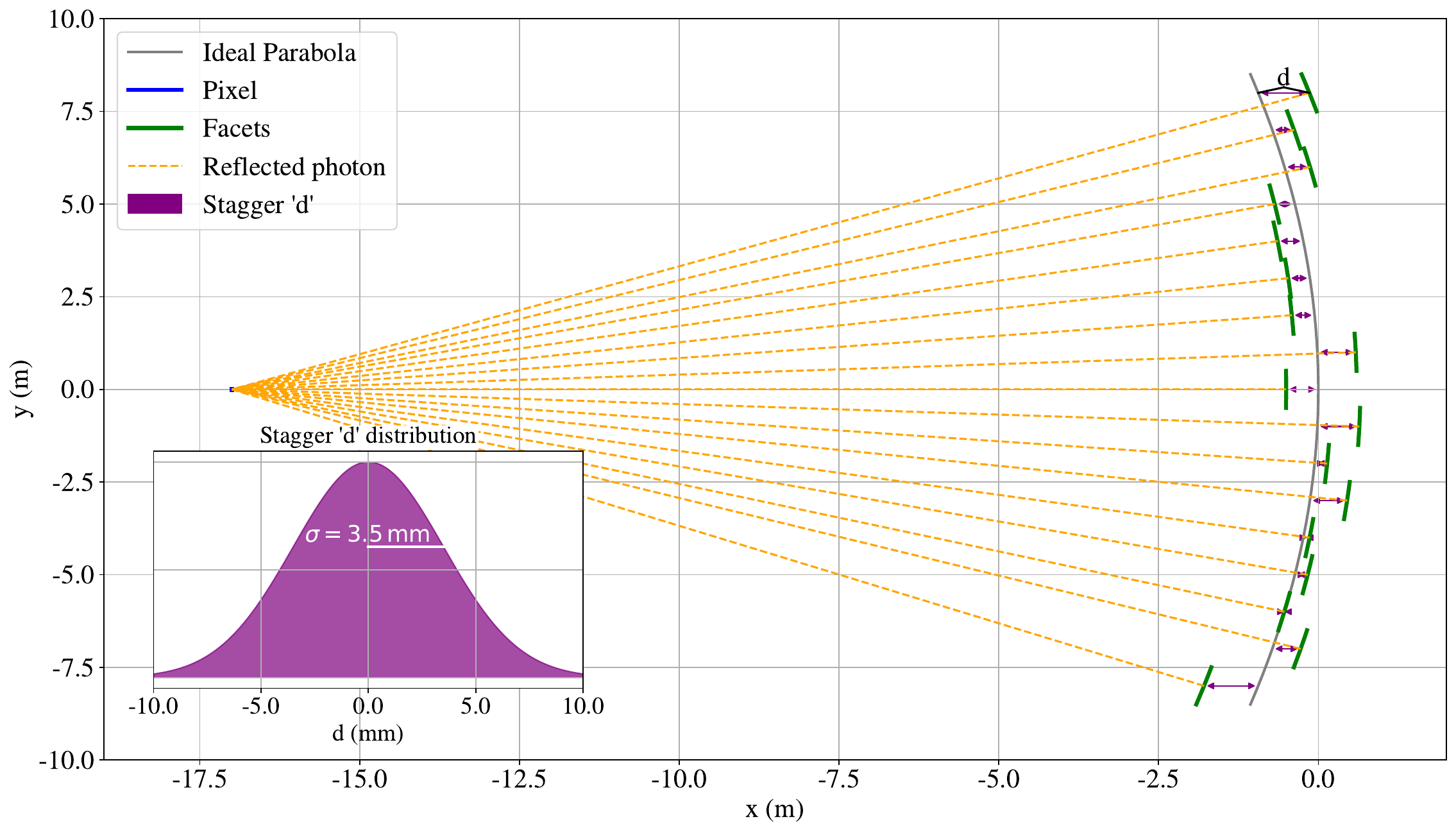}
	\caption{
          \label{fig:Stagger_illustration}
          Illustration of the random stagger of the mirror tiles caused 
          by the rough precision of the mechanical installation that is common in Cherenkov telescopes. tiles are staggered respect to the ideal paraboloid of focal length 17 m. For the second MAGIC telescope the stagger roughly follows a gaussian distribution 
          of 3.5 mm standard deviation, as illustrated in the inset figure. 
          Note that the magnitude of the stagger has been exaggerated in the main figure.}
\end{figure*}
We used Zemax (OpticsStudio) to make a ray-tracing simulation of the lens at a wavelength of 550 nm. Despite the non-optimized lens configuration, the simulation shows that all the photons fall within a spot of 3 mm diameter at the photodetector plane. The spot could in principle be made even smaller, but a complete optimization should rather take into account the expected range of wavelengths that the interferometer would operate with, and may involve an achromatic lens system.

As already pointed out in \cite{I3T}, one requires a baffle to reject light from neighboring tiles. Even if the lens and photocathode reject stray light, one may also add a cone with an entrance diameter of 30 mm and exit diameter of 3 mm between the lens and the HPD.

%

The optical PSF of each individual mirror tile is smaller than the PSF of the entire reflector, as shown in \cite{ph_m_garczarczyk}. The light spot is much smaller than a pixel for tiles near the center of the reflector. For tiles close to the edges, the light spot has a distorted elliptical shape due to the coma aberration: light is reflected on these tiles with an angle to their optical axes and spreads in the sagittal direction. Hence, pixels aligned with the central tiles will have a smaller PSF at the photodetector plane. One may use a narrower passband filter or consider adding a diaphragm before the lens to remove a large fraction of the NSB. However, a full optimization would also require considering the tracking precision of the telescope ($\pm$6 mm, following \cite{magic_upgrade2}) and goes beyond the scope of this work. In what follows, we will conservatively assume a PSF of 30 mm.

A related aspect is the precision of the alignment of the pixel and the tile. In particular, the MAGIC camera noticeably sags in the vertical plane as the telescope moves from zenith to horizon. The AMC does shift the light spot back inside the pixel, but the sag introduces an angle between the pixel pointing direction and the line between the pixel and the tile. Since the focal length is much longer than the typical sagging, the effect remains small. For a typical 30 mm sag, the misalignment is 0.1$^{\circ}$. This is less than 5\% of the angle subtended by the tile.

The specific case of the first MAGIC telescope (MAGIC-1) deserves a short discussion. Due to a design flaw, the 1 m$^2$ tiles of this telescope are installed in two parallel layers separated by around 6 cm, following a chessboard arrangement. The light coming from the front layer will obviously arrive well before the light from the back layer. In what follows, we shall ignore this situation and study only the case of the second MAGIC telescope (MAGIC-2), where there is one single layer of mirrors. For MAGIC-1, one could double the number of pixels in the butterfly 2/3 cameras and focus the light from each layer into half of the pixels. This would have hardly any effect on the conclusions we will draw: it only affects the practical implementation in MAGIC-1.

\subsection{Simulation of the photon arrival time}

We have written a Monte Carlo simulation code in Python that takes into account the general parabolic geometry of the MAGIC reflector, with its aforementioned focal length and diameter. The code calculates the photon optical path and corresponding propagation time from an initial point at the camera plane to a random position in a submirror and then to a round pixel entrance window of 30 mm diameter. 

MAGIC and LST mirror tiles have three connection points to the telescope mechanical structure. One of them is a so-called 'fixed point', which is only allowed to rotate, whereas the other two are equipped with motor actuators that enable to tilt the tile. As a consequence, gaps between adjacent tiles appear when their alignment deviates from their optical axis. The simulation allows us to calculate the impact of this gap on photon arrival times. A submirror will consist of either 1, 4 or 36 1~m$^2$ tiles, depending on the optical configuration. Each of the tiles in the submirror will be individually tilted so that light reflects on the corresponding pixel. 

In addition, the designs of both MAGIC and LST have some freedom in the distance between the fixed point and the telescope structure, because this is not critical for $\gamma$-ray observations. As a consequence, there is a few millimeter \textit{stagger} of the individual tiles with respect to the ideal parabolic shape of the reflector. The code allows to introduce a random Gaussian shift along the optical axis direction. Fig. \ref{fig:Stagger_illustration} illustrates how we simulate this random stagger: the tile is shifted by a distance d from the ideal paraboloid shape of the reflector in the x direction, i.e. along the optical axis. The stagger distance d follows a Gaussian distribution of 3.5~mm standard deviation, which has been measured to roughly correspond to the stagger in the second MAGIC telescope.

Our simulation considers submirrors centered along the vertical plane of the reflector and pixels on the same vertical plane. Mirror tiles within the submirror are simulated in positions following the curve of the reflector, i.e. the tile centers may be outside the vertical plane. This setup can be practically realized in a future Butterfly camera. The 4~m$^2$ submirror has a square shape, while the shape of the 36~m$^2$ submirror is roughly circular.

\begin{figure} 
	\begin{center}
        \includegraphics[trim=0.55cm 0.45cm 0.5cm 0.5cm, clip,width=0.475\textwidth]{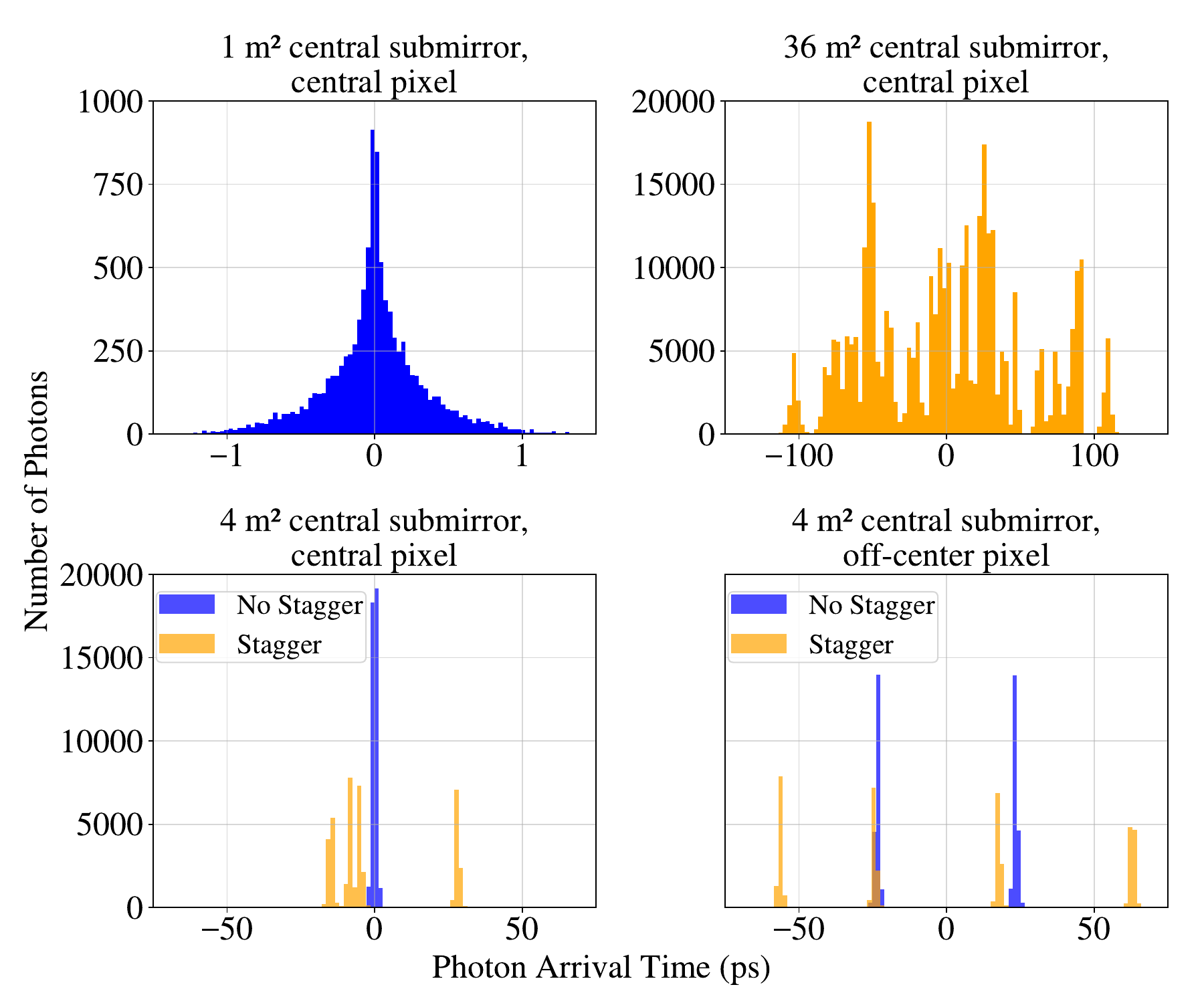}
  \end{center}
	\caption{
          \label{fig:time_distributions_all_submirrors}
          Simulated arrival time distribution, in ps, of photons arriving at a pixel. The origin of the X axis is arbitrary because we are only interested on the shape of the time distribution. Distributions for simulations with random stagger are colored orange, those without stagger are colored blue. The upper left panel corresponds to a 1~m$^2$ submirror (Butterfly 1) at the center of the reflector and a pixel at the camera center. The upper right one to a 36~m$^2$ submirror (Butterfly 3) for the same pixel and a submirror again at the center of the reflector. 
          The lower two plots corresponds to 4~m$^2$ submirrors (Butterfly 2). 
          On the left, the submirror is at the center of the reflector and the pixel at the camera center. On the right, the submirror is at the edge and the pixel is 24 cm away from the camera center.
            }
\end{figure}
Tiles have a spherical shape, so their surface only follows approximately the reflector parabolic shape. In addition, imperfections in mirror production may lead to errors in their radius up to $\sim\pm$ 20 cm \citep{ph_m_garczarczyk}. This means that even assuming that the center of the tile falls on the ideal parabolic shape, photons at the edge could be ahead or behind this shape. However, the impact of these deviations over photon paths remain well below 100~$\mu$m, so the effect can be safely neglected for all Butterfly configurations. We have also neglected the effect of imperfections on the mirror surface.

Each pixel is assumed to be a circle of 30~mm diameter. We tilt this circle so that it is perpendicular to the line connecting it to the center of the submirror. The effect of this tilt is relevant when the submirror is far from the center of the reflector. The optical PSF of the telescope follows a Gaussian distribution (with roughly 80\% of the light contained in a pixel) and simulated photons hit the pixel entrance window following that distribution.


%
\subsection{Arrival time distributions, impact in time resolution}

Here we study how the three Butterfly configurations affect the photon arrival-time distribution arriving at the pixels.


%
\subsubsection{Butterfly 1}
%

%
%
Consider the first optical configuration, with the reflector broken up into 250 submirrors. As mentioned above, we would need to install a camera with a radius of 25 cm so we will study the impact of focusing the submirrors both to the pixel near the center of the camera and to a pixel at a distance of 24 cm to the center of the camera (location of the pixel currently used for intensity interferometry observations in MAGIC). 

The upper left panel of Fig. \ref{fig:time_distributions_all_submirrors} shows the resulting arrival time distributions for a pixel at the camera center and a submirror at the center of the reflector. The distribution has a FWHM less than 1~ps. This is significantly smaller than the time resolution of any photodetector under consideration, so we can safely ignore the time spread introduced by the submirror in this Butterfly configuration. We have also simulated a submirror at the edge of the reflector and a pixel at the edge of the current MAGIC camera ($>$50 cm away from the optical axis) and the time spread introduced by the optics remains negligible. 

Note that the random tile stagger has no effect for this Butterfly configuration as the additional delay would shift the distribution, not affecting its shape.

%
\subsubsection{Butterfly 2}
%
%
Consider the second Butterfly configuration, namely a reflector broken up in 64
submirrors of 4~m$^2$. The two lower panels of Fig. \ref{fig:time_distributions_all_submirrors} show the simulated arrival time distributions for the central submirror. We compare simulations with (in orange) and without (in blue) random stagger. The left panel corresponds to a pixel at the center of the camera while the right one corresponds to a pixel 24 cm away from it. 

With no random stagger, the time distribution of the central pixel is very narrow, less than 5 ps, whereas the pixel 24 cm from the center show two distinct distributions separated by $\sim$40 ps. They are due to the different tilt in the two pairs of tiles in the submirror.


When the random stagger is added to the simulation we find four separate distributions because each tile introduces a fixed random delay to the photons. The specific delay will change from tile to tile. In other words, the time distribution will get broader than for the Butterfly 1 but its exact width will depend on how well the tiles will be aligned for that specific submirror. We obtain an average FWHM of $\sim$50 ps for the central pixel and $\sim$80 ps for a pixel 24 cm away from the camera center.

\subsubsection{Butterfly 3}
%
%
Finally, consider the third Butterfly configuration, namely a reflector broken up in only 7 submirrors of 36~m$^2$. All pixels will be located less than 10 cm from the camera center so we will only simulate photons arriving at the center of the FOV.

The upper right panel of Fig. \ref{fig:time_distributions_all_submirrors} shows the time distributions corresponding to the central submirror. The width of the distribution is due to a combination of the tilt of the tiles to focus the light to the pixel and the random stagger of the 36 tiles. We find a comparable width for submirrors centered at any distance to the optical axis. We obtain an average FWHM of $\sim$60 ps, that is, similar to the FWHM in the Butterfly 2 configuration.

\subsubsection{Impact in time resolution: summary for all configurations}

\begin{deluxetable}{lc}
\tablecaption{Time spread introduced by the three Butterfly optical configurations 
('optical time spread').\label{tab:sumary_time_distributions}}
\tablehead{
\colhead{Configuration} & \colhead{FWHM}
}
\startdata
Butterfly 1 (250 pixels) & $<$1 ps \\
Butterfly 2 (64 pixels)  & 80 ps \\
Butterfly 3 (7 pixels)   & 60 ps \\
\enddata
\end{deluxetable}
We summarize the impact of the three configurations in Tab. \ref{tab:sumary_time_distributions}. For Butterfly 2 we have actually tabulated the worst case time spread although many of the camera pixels would be in the outer rings of the camera so it will be close to the average case for the whole camera. 

In general terms, the time spread for all configurations is smaller than the time response of several of the above-mentioned photodetectors. We will add this 'optical time spread' in quadrature to the time response of the photodetector to calculate the time resolution of each Butterfly and evaluate its scientific performance.


\subsection{Other Cherenkov telescopes}

Installing a Butterfly camera in IACTs with a non-parabolic primary mirror, such as H.E.S.S., VERITAS or the CTAO MSTs, would allow to partially offset the non-isochronicity of the optics. H.E.S.S. and VERITAS do not have the same AMC capabilities of MAGIC/LSTs so the corresponding investment would be significant but the MSTs do: in their case only a software upgrade would be required.

Working with individual tiles will most probably reduce the time spread below 10-20 ps. But only a full study taking into consideration all relevant parameters of the tile arrangement (like the random stagger of the tiles) would allow to find out if the reflector could be split into submirrors larger than one tile.

\section{Expected photon detection rates}

After considering the aforementioned photodetection technologies, we will opt for a Hybrid Photodiode for the rest of this study due to its photocathode size, time resolution and QE. More specifically, we will consider the HPD module HPM-100-40 of manufacturer \citeauthor{becker_and_hickl}. This photodetection module is equipped with a Hamamatsu HPD with a photocathode of QE peaking at 45\% at 500 nm. It has a TTS of FWHM=130 ps, which roughly represents a factor 25 improvement in time resolution compared to the current PMT of MAGIC. 

Let us consider if, from the point of view of the optics, it is realistic to use this HPD model in the three MAGIC Butterfly configurations. The HPD photocathode diameter is rather small: 3~mm. It can be coupled to the above described 30~mm lens for the 1 or 4~m$^2$ configurations. The 36~m$^2$ submirror configuration needs a more optimized approach. We will assume that we can also find a practical solution for the 36 m$^2$ Butterfly configuration and we will adopt the QE and time resolution of model HPM-100-40 in our simulations.

\begin{deluxetable}{lcc}
\tablecaption{Optical and electronic parameters of the current MAGIC PMTs ("Present") and the HPDs proposed for the Butterfly camera.\label{tab:parameters}}
\tablehead{
\colhead{Parameter} & \colhead{Present} & \colhead{Butterfly}
}
\startdata
Photo-detector QE ($\alpha(\lambda_0)$)        & 0.33     & 0.45   \\
Optical efficiency ($q(\lambda_0)$)            & 0.304    & 0.304  \\
Electronic bandwidth ($b_\nu$)                 & 125 MHz  & 2.3 GHz \\
Spectral distribution ($\sigma$)               & 0.87     & 0.87   \\
Noise factor ($F$)                              & 1.15     & 1.05   \\
\enddata
\end{deluxetable}
We tabulate in Tab. \ref{tab:parameters} the relevant optical and electronic parameters of the current MAGIC optical interferometer as well as a MAGIC interferometer equipped with a Butterfly camera and the aforementioned HPDs. An improvement is also expected for the spectral distribution ($\sigma$) in the case of the Butterfly configuration, due to the lower light incidence angle into the filters, but we will be conservative and assume a value of 0.87 for the rest of the paper.  

Figure \ref{fig:Phe_rates} shows the photodetector rates that are expected for all three Butterfly configurations, using the parameters in Tab. \ref{tab:parameters} and adopting a filter with an optical passband of 2 nm (FWHM), which can be realistically applied to all configurations.

Let us see down to which magnitude we can work in single-photon mode, i.e. which stars generate a rate below 100 MHz. For Butterfly 1 (250 pixels) or Butterfly 2 (64 pixels), we could take data for essentially all stars (i.e. down to B=0$^\mathrm{m}$) whereas for Butterfly 3 we would be limited to stars dimmer than B=3$^\mathrm{m}$. This is not a major limitation.

The lowest possible detection rate is limited by either the photodetector or the NSB. This HPD model has a DCR of 400 counts per second at ambient temperature so we can safely ignore it.
NSB thus remains the limiting factor in rate: as we have said the NSB in a MAGIC telescope
corresponds to an 8.5$^\mathrm{m}$ star and is independent of the Butterfly configuration.
This would not be the case, for instance, if we selected a SiPM or a SPAD because 
one would expect a DCR of at least 300 kHz 
for a 3 mm diameter device, hence limiting the interferometer to $<$4$^\mathrm{m}$ stars.

\section{Performance}

As discussed in the previous sections, the proposed upgrades would significantly enhance the capabilities of IACTs as optical intensity interferometers. Table \ref{tab:parameters} shows how the main parameters involved in the S/N calculation would change if these modifications were implemented, either in MAGIC or LST1. The key improvements provided by the upgrade would be:

\subparagraph{Imaging capabilities in the 2-40 mas regime:} As each individual telescope would become an array of sub-mirrors, IACTs would provide access to the 1-17(23) m range for MAGIC (LST1), which corresponds to angular features in the $\sim$2-40 mas range. The dense UV coverage in all directions would be unprecedented in the optical regime, and would provide imaging capabilities for bright targets of that size. 

\subparagraph{Sensitivity boost:} When correlating different telescopes separated by longer than $\sim$50~m, UV coverage is barely affected by the modifications proposed here, as the distance between the telescopes dominates over distance between the sub-mirrors. However, the sensitivity of these correlations would be greatly enhanced, mainly due to the improvement in electronic bandwidth and quantum efficiency. 

\subparagraph{Narrow optical passband:} Stellar astrophysics generally produce much spectral lines narrower than 1~nm (expanded via thermal, rotational or turbulent broadenings). The Butterfly concept re-defines IACT's optics, allowing to employ significantly narrower optical passbands (down to 0.5 nm). Such narrow optical passbands allow to target certain scientifically relevant emission lines such as Balmer lines, which in turn opens up synergies between long-baseline (e.g. measuring the stellar photosphere of OB stars) and short-baseline observations (e.g. targeting Balmer lines to measure the large extension of decretion discs). 

A narrower passband has an additional benefit. The measured source parameters are an average over the passband, weighted with the source's spectrum (an effect referred to as 'bandwidth smearing' \citep{bandwidth_smearing}). Reducing the passband leads to less systematic uncertainty in these parameters.

\subparagraph{Beam collimation and dichroics:} The proposed setup allows to realistically collimate the light beam, which opens the possibility of using dichroics or diffuser plates to simultaneously observe several spectral channels. As discussed in section \ref{subsec:photodetectors}, the sensitivity on an intensity interferometer does not depend on the actual width of the optical passband (only on its shape), and simultaneously observing N$_{\mathrm{spectral}}$ channels improve S/N by a factor $\sqrt{\mathrm{N}_{\mathrm{spectral}}}$. In addition to the raw improvement in S/N provided, such a system would allow simultaneously imaging sources over continuum and Balmer (and other) lines. 
It should be noted that, due to the size of the expected collimated beam of IACTs, having a high N$_{\mathrm{spectral}}$ ($\sim$100) is technically challenging. A specific design exploring a feasible number of spectral channels goes beyond the scope of this paper.

\begin{figure}
  \begin{center}
      \includegraphics[width=0.48\textwidth]{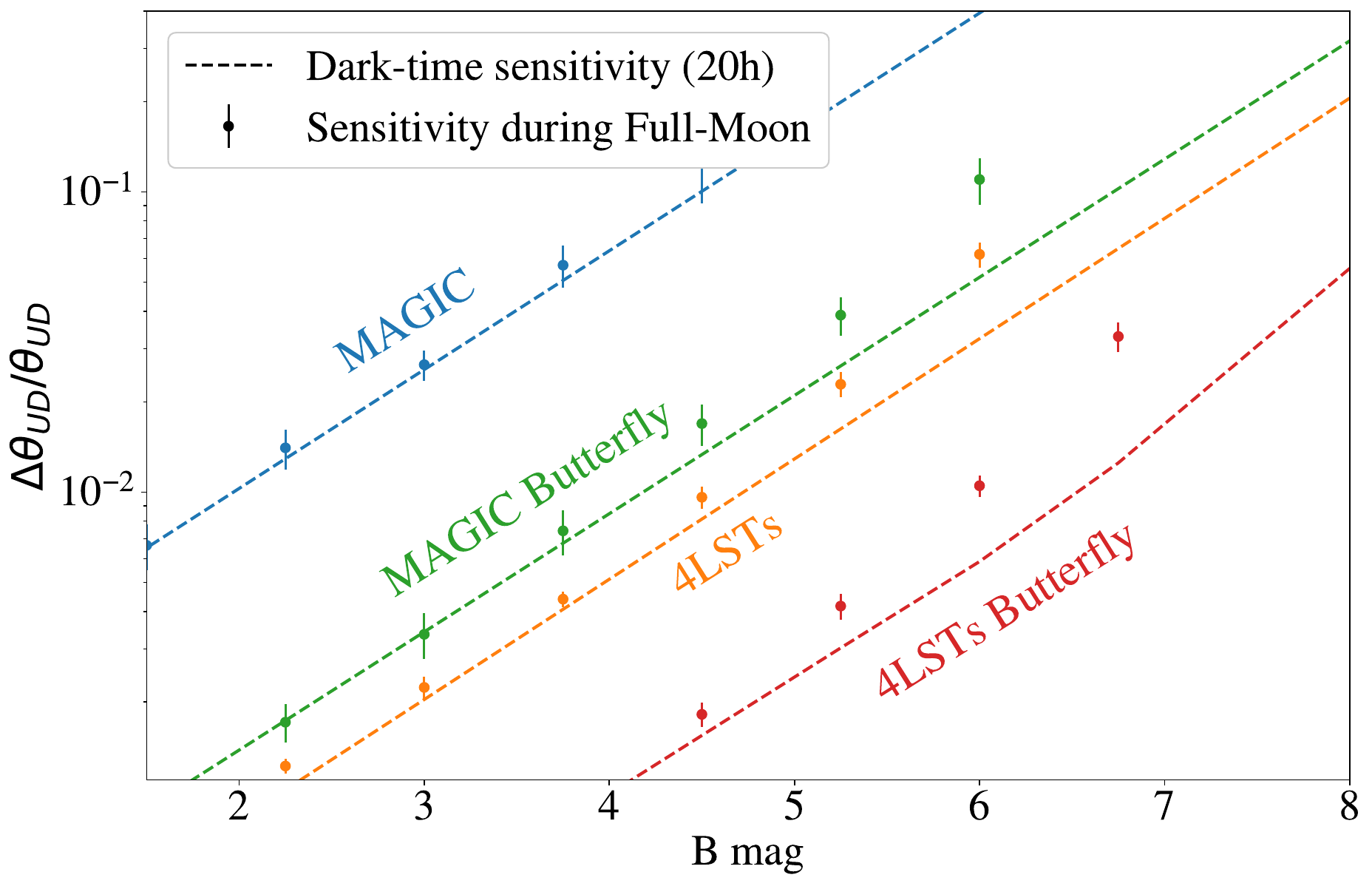}
  \end{center}
  \caption{ \label{fig:array_sensitivity}
          Relative uncertainty of measured stellar diameters as a function of their B magnitude for the current MAGIC interferometer and future 4 LST array, before and after considering the sensitivity boost achieved with the Butterfly camera. Dashed lines indicate sensitivity for 20 hour observation time during dark time, while solid points estimate the sensitivity for very high NSB illumination levels, namely 10 times Dark NSB (NSB in MAGIC typically ranges between 5 and 30  times Dark NSB during the Full Moon nights, see \cite{moon}). The simulated stellar position and diameter is that of $\gamma$~Crv. Note that the impact of the NSB on the LSTs is overestimated, as we assume that the optical PSF of MAGIC and the LSTs are identical.}
\end{figure}

Taking a relatively simple implementation with N$_{\mathrm{spectral}}$ = 1 and the parameters described in Tab. \ref{tab:parameters}, we are able to estimate the impact of the Butterfly camera on the high-level performance of both MAGIC and the planned 4 LST array. Fig. \ref{fig:array_sensitivity} shows the impact of the proposed setup on the measurement of a star with the same diameter (0.72 mas) and declination as the bright calibrator star $\gamma$~Crv but with varying magnitude. Equipping MAGIC with any of the Butterfly configurations results in an increase of sensitivity of roughly a factor 6 or 2 magnitudes. For the 4 LSTs the increase is smaller because the current PMTs already reach a peak QE around 45\% but still significant: 1.5 magnitudes. For these telescopes one observes a noticeable degradation of performance during Full Moon because the Butterfly allows reaching stars weaker than 5$^m$, for which the background light from the bright Moon is comparable to the light from the star. 

\newpage
\section{Scientific prospects}
\begin{figure}
    \begin{center}
         \includegraphics[width=0.5\textwidth]{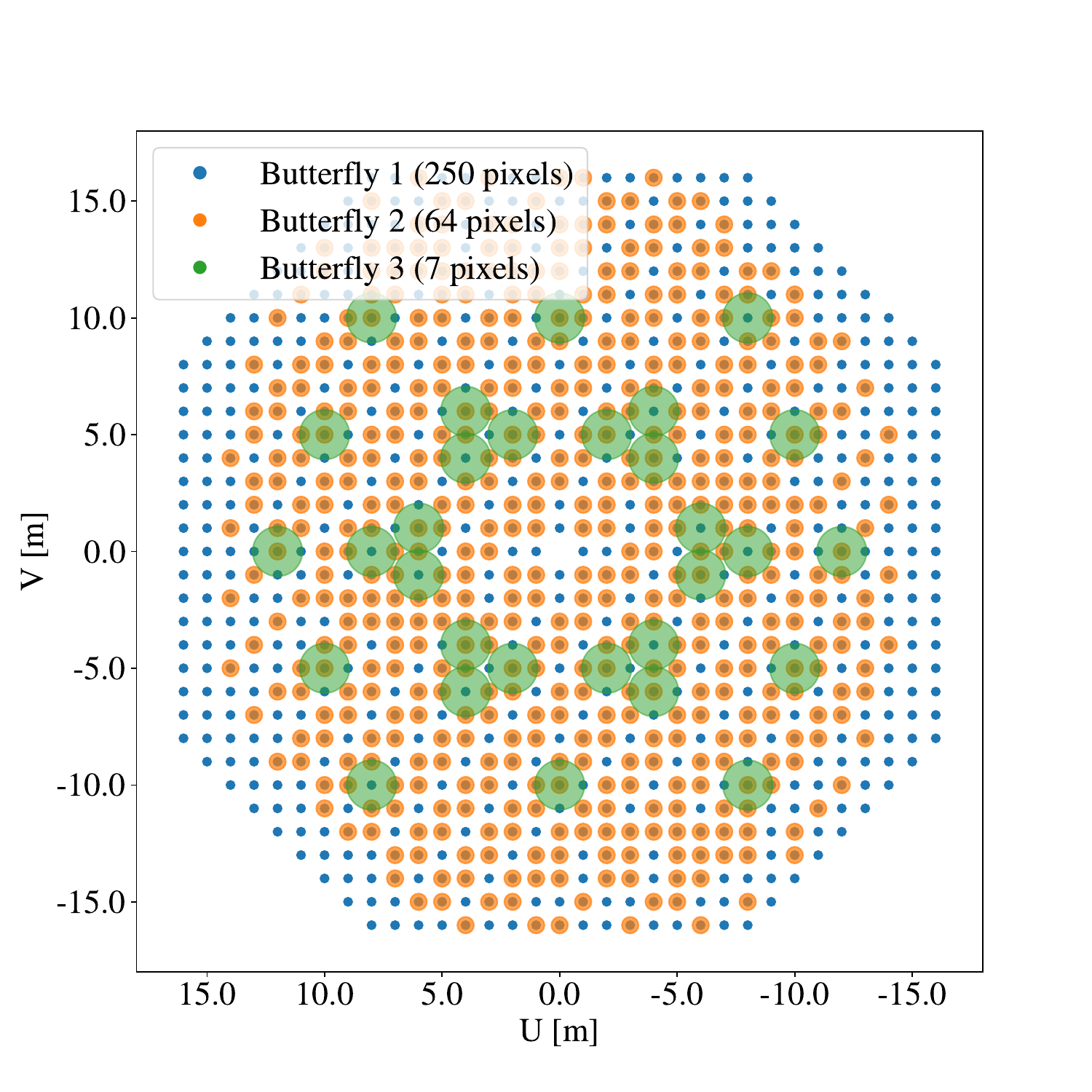}
    \end{center}
    \caption{ \label{fig:uv_coverage}
          Instantaneous UV coverage of the three proposed Butterfly configurations for a MAGIC telescope. Short baselines will be observed by multiple pixel pairs. For the Butterfly 1 configuration the maximum number of multiplicities is 229 at the shortest baselines, while for Butterfly 2 and Butterfly 3 configurations the maximum multiplicities is 47 and 2 respectively.}
\end{figure}

\begin{figure}
    \begin{center}
         \includegraphics[width=0.49\textwidth]{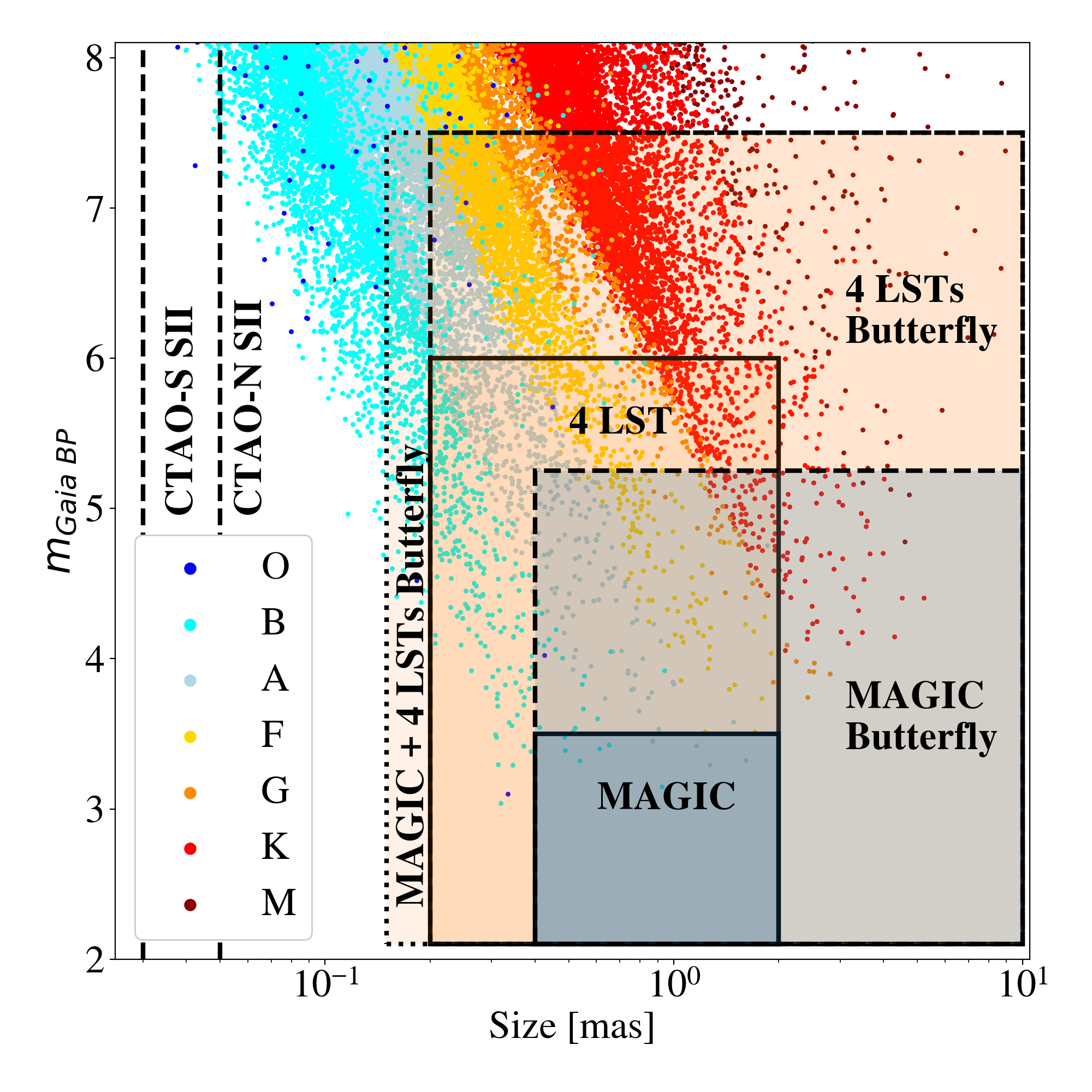}
    \end{center}
    \caption{\label{fig:gaia candidates} Stellar magnitude measured with Gaia's blue passband filter ($G_{BP}$) vs angular diameter of bright stars in the Gaia DR3 catalog \citep{gaia_dr3} with estimated diameter $0.01$ mas $< \Theta < ~10$ mas and $m<8$. In order to obtain the angular diameter for each star, we have used the stellar parameters from \cite{gais_stellar_parameters}. The limiting magnitude (corresponding to a 3\% relative uncertainty over 20 h)  and angular resolution is shown for different arrays of IACTs, with (dashed contours) and without (solid contours) the Butterfly configuration. The maximum angular resolution of the North and South arrays of CTAO is shown as dashed vertical lines.
          }
\end{figure}

\begin{figure*}
    \begin{center}
         \includegraphics[width=0.99\textwidth]{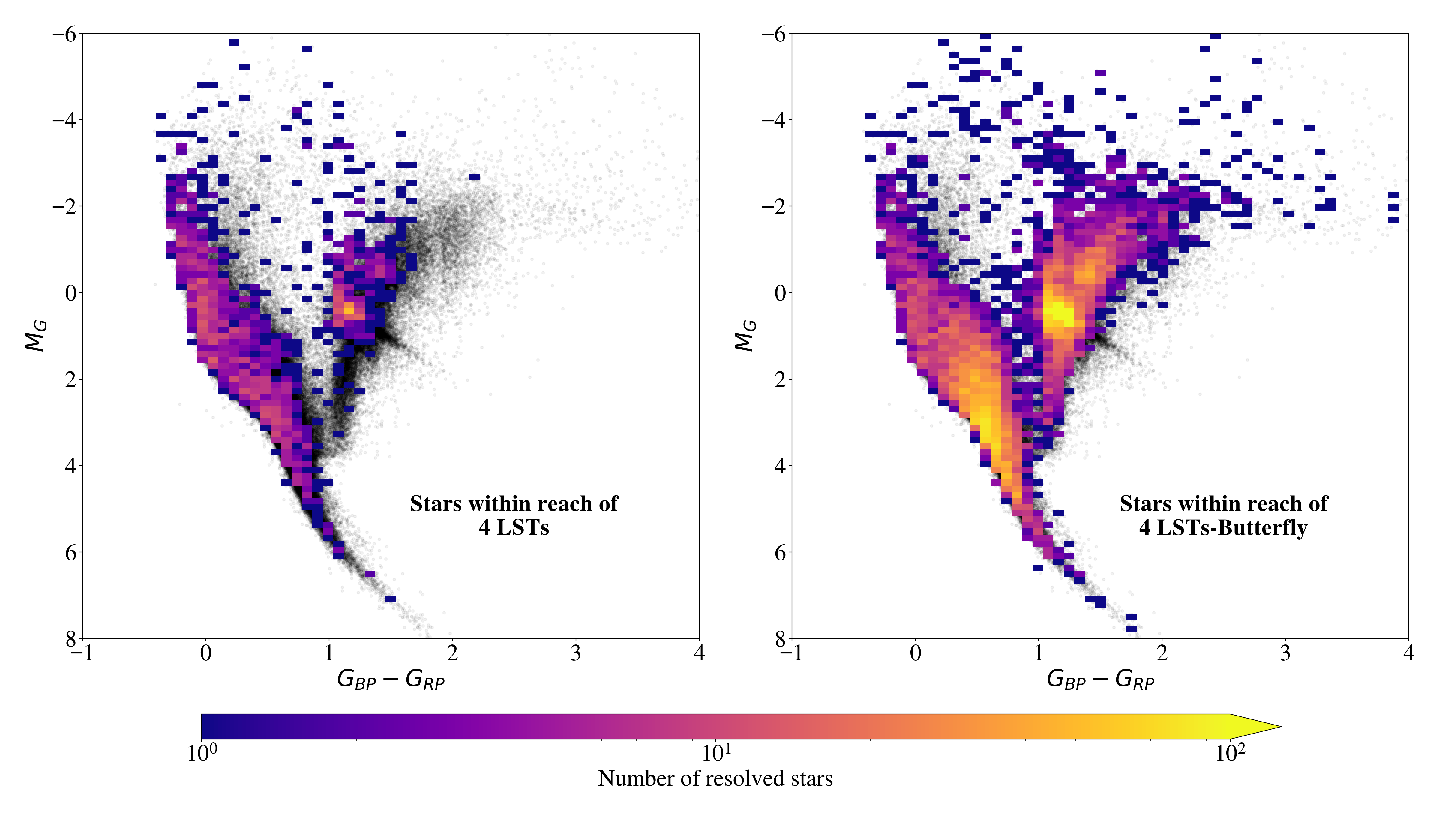}
    \end{center}
    \caption{\label{fig:hr_diagrams}Color-magnitude diagrams of stars from the Gaia DR3 catalog \citep{gaia_dr3} with $m_{BP} < 10$. The stars are shown as gray points in both figures. A color map is shown on top of the stars indicating the number of stars that can be resolved with the 4 LSTs array (left figure) and 4 LSTs equipped with a Butterfly camera (right figure), accounting for both sensitivity and angular resolution.
          }
\end{figure*}

It should be clear from Fig. \ref{fig:array_sensitivity} that a MAGIC or LST Butterfly array would bring about a very significant {\bf improvement in sensitivity} over the full range of UV. This would enlarge the populations within reach and has a strong impact on all science cases that have been proposed for IACTs and specifically for CTAO \citep{cta}:
\begin{itemize}
    \item Systematic measurements of stellar diameters for all spectral types, with precision down to 1\%.
    \item Focus on early types, like blue supergiants and related stars.
    \item Orbital parameters of binary systems.
    \item Interacting binaries, where one could study mutual irradiation, tidal distortion, limb darkening, rotational distortion, gravity darkening, and oscillations.
\end{itemize}

To better illustrate this point, Fig. \ref{fig:gaia candidates} presents the stellar populations observable by different IACT arrays. It showcases projections for the future MAGIC+4LSTs array, both independently and when equipped with Butterfly cameras. The improved sensitivity from the Butterfly cameras is significant, expanding the number of stars detectable within the MAGIC+4LSTs array's limiting magnitude by a factor of nine. Crucially, the Butterfly LSTs facilitate the observation of a substantial population of B stars, particularly when integrated with the two MAGIC telescopes to provide longer baselines.

Combining the 4 LSTs with the other telescopes of the CTAO arrays would allow reaching stars with size significantly smaller than 100 $\mu$as. 

This would also be possible by combining the 4 LSTs in CTAO-North with existing optical telescopes at ORM in La Palma, and more specifically with the Gran Telescopio Canarias (GTC, 10.4 m in diameter). Adding an HPD-based one pixel camera of small dimension to an optical telescope is conceptually simple.

The benefit of the Butterfly cameras is also evident in Fig. \ref{fig:hr_diagrams}. Here we show color-luminosity diagrams of the same stars and we compare the stars whose diameters we can resolve with the current LSTs with those we could resolve with the same telescopes equipped with Butterfly cameras.  

A narrower optical passband allows observing {\bf broad emission lines}. That would prove instrumental in other proposed scientific goals where the emission region is bright in these lines:
\begin{itemize}
    \item Circumstellar disks.
    \item Winds from hot stars.
    \item Wolf–Rayet stars and their environments.
    \item Luminous blue variables.
\end{itemize}

And, even more importantly, the Butterfly array adds new science cases. Figure \ref{fig:uv_coverage} shows the UV coverage of a single MAGIC Butterfly telescope for each configuration. Due to this very dense instantaneous UV coverage, single MAGIC-Butterfly observations would be ideal to study the shape of any object with {\bf angular scales ranging between 2 and 40 mas}. Especially for Butterfly 1 the UV coverage is so dense that one expects 'pseudo-images' of these objects. 

\cite{I3T} already pointed to a high-impact science case: imaging the surface of Betelgeuse as it evolves in time.
 
In this section, we will go deeper in two other scientific goals strongly benefiting from such an interferometer, namely measuring the shape of fast rotating stars and their decretion disks, and measuring the expanding shells of Novae/SNe outbursts. We will make specific predictions for these objects and our instrumental setup.

\subparagraph{Fast rotating stars and decretion disks:} The broad UV coverage provided by the MAGIC Butterfly is particularly well suited to directly measure some of the key properties of these stars: by measuring their oblateness we will infer their rotation, which has been identified as an important parameter affecting their evolution, perhaps as important as mass and metallicity \citep{Maeder}. After the initial success of amplitude interferometry observations of Altair, Achernar, Regulus, Alderamin, Vega, Rasalhague and $\beta$ Cas, a list of other fast rotating targets have been identified as suitable for similar studies \citep{vanBelle_catalog}. Even though most of the fast rotating stars shown in this catalog have angular diameters below 2 mas, hence better suited for multi-telescope array observations, we identified 7 candidates suitable to be imaged by a single MAGIC-Butterfly. This number of candidates increases if narrow-band observations of decretion disks are taken into account. On the other hand, an array of 4LSTs equipped with a Butterfly camera could be used to study fast rotating stars with angular diameters below $\sim$ 2 mas.

\begin{figure}
    \begin{center}
    \includegraphics[width=0.5\textwidth]{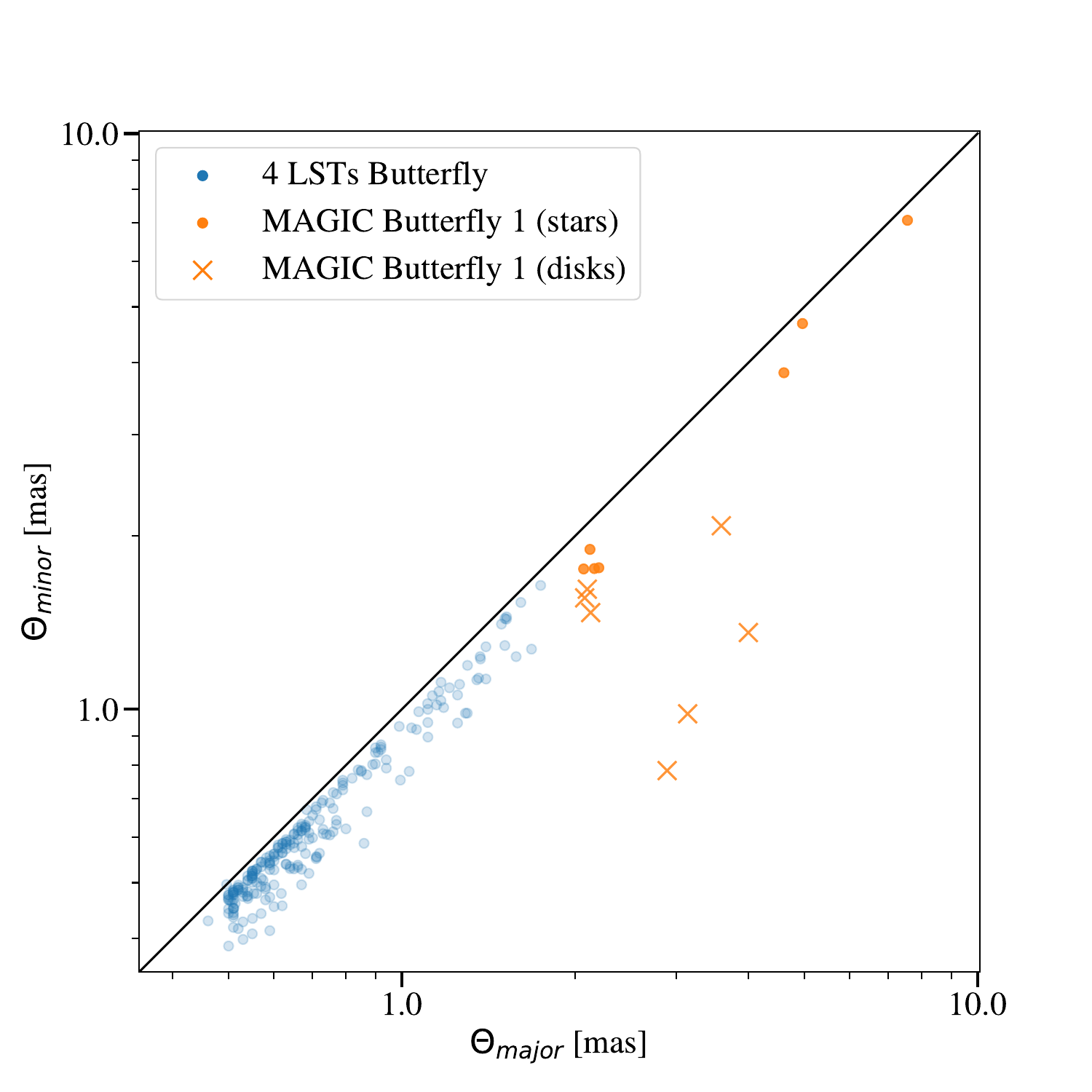}
    \end{center}
    \caption{\label{fig:fast rotators} Simulations of observations with the MAGIC Butterfly 1 configuration of fast rotating stars (orange circles) and known decretion disks (orange crosses).  Simulations of observations with 4 LSTs equipped with a Butterfly camera of fast rotating stars are also shown (blue circles). The solid line shows $\Theta_{\mathrm{minor}} = \Theta_{\mathrm{major}}$  i.e no oblateness. The distance to the solid line is proportional to the oblateness of each simulated star. The uncertainties both for the major and minor angular diameter of these points are bellow the 2\% level, which is smaller than the marker size and low enough to discard non-oblate models for any simulated star.   
    }
\end{figure}

Fast rotating stars, especially those showing a decretion disk, are known to be complex systems showing variability over periods shorter than days, strong emission and absorption lines (also variable with time) and multiplicity \citep{rivinius_classical_2013}. Simulating all these effects goes beyond the scope of this study and, therefore, we will use a very simple model in order to get an estimate of the number of stars that could be studied with different arrays of IACTs equipped with the Butterfly camera. For each star, we have simulated emission coming from a uniform ellipse, taking the values from \cite{vanBelle_catalog} for the oblateness and the major angular diameter, and a random value between 0 and 180 degrees for the position angle (PA) of the stellar equator. Figure \ref{fig:fast rotators} shows simulations of observations on most of the fast rotating stars in \cite{vanBelle_catalog}. For each star we have simulated short 2.5 hours observations under high Moon illumination levels and different interferometers. For the MAGIC Butterfly 1, we selected stars from \cite{vanBelle_catalog} with the following constraints: $DEC \geq -35 ~\mathrm{deg}$, $V_{\mathrm{mag}} \leq 4$ and $2 ~\mathrm{mas} \leq \theta < 40 ~\mathrm{mas}$. We also evaluated the potential of using 4 LSTs equipped with a Butterfly 3, selecting stars with the following constraints: $DEC \geq -35 ~\mathrm{deg}$, $V_{\mathrm{mag}} \leq 6$ and $0.3 ~\mathrm{mas} < \theta < 2 ~\mathrm{mas}$. We also performed simulations to explore the detectability of decretion disks (assuming again a uniform ellipse model): using the MAGIC Butterfly 1 equipped with a $H_{\beta}$ narrow-band filter, we simulated 7 different Be star's decretion disk measured by \cite{CHARA_be}.

Simulations show that the 7 stellar and 7 decretion disks candidates selected for MAGIC Butterfly 1 are clearly detected in 2.5 hours, with uncertainties reaching the 1\% level for all the simulated parameters (i.e the oblateness, major angular diameter and PA of the equator). In the case of observations with the 4LSTs equipped with the Butterfly 3, we have a very similar result for a much larger sample of 226 candidates, with uncertainties reaching the 1\% level again. The studied sample is not complete, as a broader sample of fast rotating Be stars presenting active decretion disks of angular scales $> 2~\mathrm{mas}$ could be studied with the MAGIC Butterfly 1. In addition, the list of fast rotating stars presented in \cite{vanBelle_catalog} only covers stars with equatorial diameters of $\Theta > 0.45~\mathrm{mas}$, while the angular resolution in the Blue band for the 4LSTs array will reach angular resolutions of $\sim 0.25~\mathrm{mas}$ with baselines as large as $180$ m.

\subparagraph{Novae:} Resolving the morphological evolution of novae ejecta at their earliest times after the outburst is key to trace the evolution of the ejecta and resolve structures that maybe be too low-density to be detected at late times, allowing to constrain the mass ejection processes and origin of shocks with high precision \citep{Classical_novae_review}. As it expands, the ejecta grow to an angular size of: 

\begin{equation}
    \Theta = 1.15 ~\mathrm{mas} \left( \frac{d}{1~\mathrm{kpc}} \right)^{-1} \left( \frac{v_{ej}}{1000 ~\mathrm{km~s^{-1}}} \right) \left( \frac{t}{days} \right)
\end{equation}

where $t$ is the time since the start of the expansion, $v_{ej}$ is the velocity of the ejecta and $d$ is the source distance \citep{Classical_novae_review}. Following this equation, typical novae will have angular sizes below 1 mas during the first night after the eruption and will grow to sizes between 2 and 40 mas during the following weeks or even months.

Optical and infrared interferometers have observed novae for some decades, Nova Cyg 1992 was the first nova resolved by optical interferometry \citep{Nova_Cyg_1992}, followed by V838 Mon \citep{V838_Mon} and Nova Aql 2005 \citep{Nova_Aql2005}, but it was not until RS Oph 2006 that the expansion rate was measured for the first time, only 5.5 days after the outburst \citep{RSOph_2007, RSOph_2007_Monnier}, followed by observations of V1280 Sco in 2008 \citep{V1280_Sco} and T Pyxidis in 2011 \citep{T_Pyxidis}. More recently, the initial fireball phase of a nova was studied for Nova Delphini 2013, observed as early as within the first 24 hours after the outburst \citep{Nova_Del_2013}. 

Temporal evolution and structure information when available proved to be very valuable in order to study the physical processes operating in the vicinity of the central remnant early after the outburst or even during the earliest fireball phase. Nevertheless, the limited UV-coverage of the observations made structure studies over multiple days very challenging. Observations with the MAGIC Butterfly configurations 1 and 2 would have an unprecedented UV-coverage and, combined with the possibility to observe at narrow bands without degrading its SNR, would provide excellent temporal and structure information during the first days after the outburst.

\begin{figure}
    \begin{center}
         \includegraphics[width=0.48\textwidth]{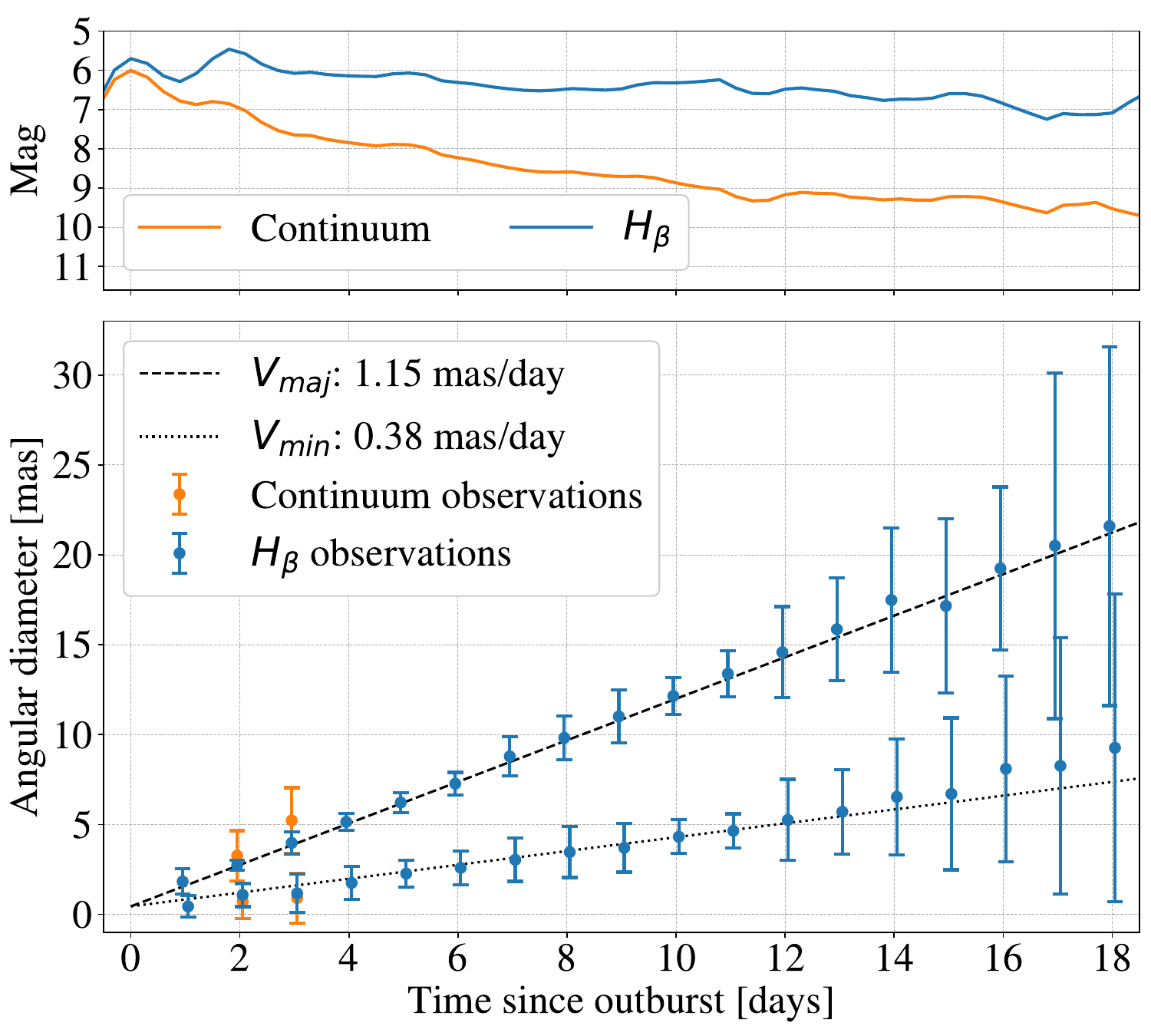}
    \end{center}
    \caption{\label{fig:nova evolution}
          Simulations of a nova located at $1.5~\mathrm{kpc}$ distance with a peak of magnitude 6 in the Visible band. On the upper plot the lightcurve is shown for both the continuum (orange line) and the equivalent lightcurve for observations carried out in the $H_\beta$ line (blue line). The reconstructed major and minor axis of the expanding shell is shown in the lower plot with the error of each measurement. Simulated observations with the MAGIC Butterfly 1 are shown in orange for continuum observations and in blue for observations with a narrow $H_\beta$ filter. The dashed black line shows the expansion rate of the major axis while the dotted line shows the expansion rate of the minor axis. Simulations for $>8^\mathrm{m}$ are not shown here due to the low SNR expected for five hours observations of $>8^\mathrm{m}$ objects.}
\end{figure}

 In Figure \ref{fig:nova evolution} we show simulations of a nova outburst located at $1.5~\mathrm{kpc}$, the size of the nova at the time of the outburst was assumed to be $5*10^{11}~\mathrm{m}$ while the major and minor axis expansion velocities were assumed to be $1500~\mathrm{km~s^{-1}}$ and $500~\mathrm{km~s^{-1}}$. We simulated two different lightcurves for the nova, one for observations in the continuum and a different one for narrow band observations in the $H_\beta$ emission line. For the continuum lightcurve we selected a peak magnitude of 6, assuming a similar brightness evolution as the one from Nova RS Oph 2021 (public data taken from AAVSO). In order to compute the brightness of the $H_\beta$ emission line we used the line normalized flux with respect to the continuum, again relying on public data from AAVSO for the same Nova RS Oph 2021. Five hours observations under dark conditions with the MAGIC Butterfly 1 were simulated for each day, centered at $500~\mathrm{nm}$ for the continuum and at $486~\mathrm{nm}$ for the $H_\beta$ observations.

Simulations in this case show that the asymmetry in the expansion rate would be resolved, with uncertainties of $\sim10\%$ for both major and minor expansion rates in the case of observations carried out in the continuum, while observations in the $H_{\beta}$ line would improve the resolution considerably, producing uncertainties on the $2\%$ level. As the only difference between the two simulated observations is the magnitude, with a faster decay in the continuum, we conclude that the main limitation during novae observations would be the magnitude of the observed nova.

The rate of novae as bright or brighter than the one simulated here (peak $<6^\mathrm{m}$) is expected to be one per year on average \citep{Shafter_2017}. In addition, the choice of the assumed lightcurve was conservative, as it corresponds to a very fast decay nova with $t_2 < 10~\mathrm{d}$ being $t_2$ the time for the nova to decay 2 magnitudes \citep{Observations_of_novae}, while novae of peak $>5^\mathrm{m}$ typically have larger $t_2$ values \citep{Observations_of_novae}.

\subparagraph{Supernovae:} Similarly to what was shown for novae, a galactic supernova structure could be resolved in the earliest time for sizes between 2 and 40 mas with the MAGIC Butterfly. As the main limitation of such observations would still be the magnitude of the supernovae, we would be sensitive only to galactic or close-by bright extragalactic SNe within a volume of $\sim100~\mathrm{kpc}$.

\section{Conclusions}

The Intensity Interferometry technique, initially pioneered by H. Brown and R. Q. Twiss with the Narrabri Stellar Intensity Interferometer, was largely superseded by amplitude interferometers, which were both theoretically and empirically shown to achieve better sensitivities \citep{HB1974}. Nevertheless, SII is currently undergoing a significant revival. This is mainly due to the emergence of large optical telescope systems in IACTs, which are employed by collaborations like MAGIC, HESS, and VERITAS, and are outfitted with fast (nanosecond), high-quantum efficiency photodetectors. Further substantial improvements in sensitivity are anticipated through technological advancements, specifically the development of picosecond synchronization capabilities and the adoption of photodetectors enabling single-photon detection with TTS on the order of tens of picoseconds.

While exploring the use of IACTs as optical interferometers, researchers have proposed solutions to extend their UV coverage. By treating individual tiles of an IACT's reflecting dish as independent telescopes and correlating their star signals, these instruments could achieve baselines in the 1-20 meter range \citep{I3T}. Specific implementations have been proposed capable of applying narrow-band optical filters with improved timing \citep{nolan_pupil}, although they do not go into the specifics of the photodetector and the optical setup lacks the capability of achieving a time resolution around 100 ps.

This study demonstrates, for the first time, that IACTs are not only technically suitable for implementing the I3T concept, as shown in \citep{spie2022}, but also capable of utilizing sub-nm optical passbands and existing photo-detection technologies with TTS in the 100 ps range using off-the-shelf components.

Three implementations, increasing in cost and complexity, are proposed here using the MAGIC telescopes as examples. The Butterfly 3 camera is the smallest camera and, therefore, easiest to implement for a reduced cost. Even if it only provides moderate access to short baselines in the UV plane (as shown in Fig. \ref{fig:uv_coverage}), it is enough to dramatically boost sensitivity due to the improvement in bandwidth. The Butterfly 1 camera not only provides such a boost in sensitivity, but also provides unprecedented UV coverage in the 1-17 m range. Such a camera would provide imaging capabilities to bright optical sources with a resolution bridging between long-baseline interferometers and space optical telescopes (e.g.  HST/WHT). As proposed by \cite{I3T}, imaging the surface evolution of Betelgeuse would be within the reach of such a telescope.

Among the different optical designs of IACTs, there are those that slightly degrade the isochronicity of the optical axis to improve off-axis performance, emplying modifications such as Davies-Cotton design \citep{DaviesCotton}. IACTs such as VERITAS, H.E.S.S. or the future Medium Sized Telescopes (MST) from CTAO use this optical design. This design was identified in the past as a key limitation of these telescopes as intensity interferometers, as the lack of isochronicity added an irreducible limitation to the bandwidth of the correlation \citep{Dravins}. The Butterfly camera removes this limitation dividing Davies-Cotton primary mirrors into smaller sections, each with dramatically better isochronicity.

These findings mean IACTs are technically suitable to improve very significantly current sensitivity limits. The dramatic improvement of IACT sensitivity as the intensity interferometers shown (see Fig. \ref{fig:array_sensitivity}) would affect every science case ever proposed for these systems \citep{Dravins}. Figures \ref{fig:gaia candidates} and \ref{fig:hr_diagrams} show that the number of objects we can resolve increases dramatically when we equip the MAGIC or LST telescopes with Butterfly cameras.
 
In addition, we explored specific science cases that would benefit from an optical interferomers sensitive to baselines within the 1-20 m range (angular scales ranging between 2-40 mas). A single MAGIC telescope equipped with the Butterfly 1 camera is capable of detecting the oblateness of 7 OB fast rotators for the first time in the Blue band (3 of them never studied before at any other wavelength) as well as imaging decretion disks in just 2.5 hours. If such a sensitivity boost was provided to a system of 4 LSTs the impact on the population of OB fast rotators imaged would be dramatic, increasing by a factor 20, as shown in Fig. \ref{fig:fast rotators}. Measuring the shape during the earlier stages of novae may also be possible with such instruments. By employing a pair of MAGIC telescopes equipped with the Butterfly 1 camera, we would be able to accurately measure the velocity and shape of the expanding shell of a nova with a peak of 6 V mag over the first $\sim$ 10-15 days from the outburst (depending, of course, on the expansion velocity and distance). 

In the revival of the SII technique that we are currently living in, IACTs are not the only optical telescopes identified to have great scientific potential as intensity interferometers. Classical large optical observatories may reach unprecedented sensitivity and resolution as intensity interferometers if they are able to exploit their excellent optics to separate the optical beam into N bands (where N could be a very large number). This would be technically feasible if large arrays of ultra-fast photodetectors were employed in cameras analog to current spectrograph and has already been demonstrated in the laboratory for a 5 spectral channel intensity interferometry by \cite{Multiplexing_SII}.

Reaching both sub-nm optical passbands and 100 ps scale bandwidths will be critical to integrate IACTs into larger intensity interferometry arrays, also including classical optical observatories \citep{sea_2022_hassan}. This possibility adds enormous upgrade potential and scientific prospects to the future CTAO. CTAO-North in the ORM will be surrounded by the GTC, the William Herschel Telescope (WHT, 4.2 m in diameter) and Telescopio Nazionale Galileo (TNG, 3.6 m in diameter). And, looking for longer baselines with sensitivities and resolution allowing extragalactic intensity interferometry, one could envisage CTAO-South array in Paranal (Chile) operating jointly with both ESO's Very Large Telescope (VLT, 4 telescopes 8.2 m in diameter each) and the Extremely Large Telescope (ELT, 39 m in diameter).


\section*{Acknowledgments}

Funded/Co-funded by the European Union (ERC, MicroStars, 101076533). However, views and opinions expressed are those of the authors alone and do not necessarily reflect those of the European Union or the European Research Council. Neither the European Union nor the granting authority can be held responsible for them.

We acknowledge many fruitful discussions with our colleagues of the MAGIC and LST collaboration, in particular those in the interferometry team. We also acknowledge an anonymous referee for his/her very useful comments.

J.C. is grateful to the Swiss Boninchi Foundation and T. Montaruli for funding a research stay at the University of Geneva, where he experimented with the use of single-phe photodetectors and correlators for interferometry. 


%

\vspace{5mm}
\facilities{MAGIC, LST-1}






\bibliographystyle{aasjournal}
\bibliography{bibliography}{}



\end{document}